\documentclass[prx,amsmath,amssymb, notitlepage, twocolumn,
nofootinbib,
superscriptaddress,
longbibliography
]{revtex4-2}

\usepackage{amsmath}
\usepackage{tabularx,graphicx}
\usepackage{epstopdf}
\usepackage{MnSymbol}
\usepackage{graphicx}
\usepackage{latexsym}
\usepackage{bbm}
\usepackage{amssymb}
\usepackage{amsthm}
\usepackage{color, colortbl}
\usepackage{psfrag}
\usepackage{bbm}
\usepackage{bm}
\usepackage{titlesec}
\usepackage{dsfont}
\usepackage{feynmp}
\usepackage{slashed}
\usepackage{multirow}
\usepackage{url}

\newcommand{\beq}{\begin{eqnarray}}
	\newcommand{\eeq}{\end{eqnarray}}

\newcommand{\tr}{{\rm tr}}

\newcommand{\bsp}{\begin{split}}
	\newcommand{\esp}{\end{split}}
\newcommand{\const}{{\rm const}}

\newcommand{\hc}{{\rm h.c.}}

\newcommand{\ie}{{i.e., }}
\newcommand{\eg}{{e.g., }}

\usepackage{color}
\definecolor{darkblue}{rgb}{0.,0.,0.4}
\definecolor{darkred}{rgb}{0.5,0.,0.}
\definecolor{BlueViolet}{RGB}{138,43,226}
\definecolor{SkyBlue}{RGB}{30,144,255}
\definecolor{DarkGreen}{RGB}{0,100,0}
\usepackage[pdftex,colorlinks=true,linkcolor=darkblue,citecolor=blue,urlcolor=darkred]{hyperref}
\usepackage[normalem]{ulem}

\renewcommand{\vec}[1]{\bm{#1}}

\newtheorem{theorem}{Theorem}

\begin{document}
	\title{Lieb-Schultz-Mattis Theorem with Long-Range Interactions}
	
	\author{Ruochen Ma}
	\affiliation{Kadanoff Center for Theoretical Physics, University of Chicago, Chicago, Illinois 60637, USA}

\begin{abstract}

We prove the Lieb-Schultz-Mattis theorem in $d$-dimensional spin systems exhibiting $SO(3)$ spin rotation and lattice translation symmetries in the presence of $k-$local interactions decaying as $\sim 1/r^\alpha$ with distance $r$. Two types of Hamiltonians are considered: Type I comprises long-range spin-spin couplings, while Type II features long-range couplings between $SO(3)$ symmetric local operators. For spin-$\frac{1}{2}$ systems, it is shown that Type I cannot have a unique symmetric ground state with a nonzero excitation gap when the interaction decays sufficiently fast, \ie when $\alpha>\max(3d,4d-2)$. For Type II, the condition becomes $\alpha>\max(3d-1,4d-3)$. In $1d$, this ingappability condition is improved to $\alpha>2$ for Type I and $\alpha>0$ for Type II by examining the energy of a state with a uniform $2\pi$ twist. Notably, in $2d$, a Type II Hamiltonian with van der Waals interaction is subject to the constraint of the theorem.

\end{abstract}

\maketitle

The spectral gap is a central quantity of a quantum many-body system, which determines the correlation and entanglement properties in the ground state \cite{2007arealaw,2010Hastingsn,2004Hastings,2006HK,20201darealaw}. Regarding the spectral gap, a kinematic constraint is the Lieb-Schultz-Mattis (LSM) Theorem \cite{lieb1961two}. It states that for a translationally invariant spin chain with $SO(3)$ spin rotation symmetry and a spin-$\frac{1}{2}$ moment per site, a unique gapped symmetric ground state is forbidden. The theorem, together with its topological argument \cite{2000Oshikawa,2017Cho} and higher-dimensional generalizations \cite{2004Hastings,2007multidimensional,affleck1988spin}, features the fundamental role played by symmetry and topology in quantum many-body physics.

However, most discussions of the LSM Theorem in current literature focus on systems with only finite-range interactions (with a few exceptions, see Ref. \cite{2021Tada}). Recently, it has been realized that long-range interactions, \ie power-law decaying as $\sim 1/r^\alpha$ in distance $r$, which are ubiquitous in many-body systems, can lead to qualitatively new physics \cite{2013quasilbr,2016Toplong,2017arealaw,2020onedreview,2023gong,2014GongG,2021Kuwahara}. Thus, it is important to ask whether and to what extent the LSM constraint holds in the presence of such interactions. In this work, we address this issue, where the main results are summarized in Theorems \ref{thm:1d} and \ref{thm:higherd}. We first define the system in $1d$ and demonstrate the existence of a low-lying state with vanishing energy above the ground state following the original proof \cite{lieb1961two,2022Tasaki}. Then, to generalize the discussion to higher dimensions, we utilize the powerful machinery of quasi-adiabatic evolution \cite{2005quasiadiabatic,2004Hastings,2010Hastingsn} to construct a low-energy state with a twisted boundary condition, which is then shown to be orthogonal to the original ground state due to the fact that each site has a spin-$\frac{1}{2}$ moment. In this reasoning, a crucial ingredient is the recent development of the Lieb-Robinson (LR) bound in power-law interacting systems \cite{2019LRbound,2021LRbound,2020hierarchy,2015Feig,2020LR2,2023LRreview,2018Else}.

\emph{Systems in $1d$.} Consider a $1d$ chain with length $L$ and periodic boundary condition. In each unit cell, there is a spin-$\frac{1}{2}$ moment. The Hamiltonian is invariant under lattice translation and an $SO(3)$ global spin rotation symmetry. Throughout this work, we will consider two types of Hamiltonians. Type I consists of two-body antiferromagnetic interactions, with strengths decaying according to a power $\alpha$ in distance,
\begin{equation}
    H_I  = \sum_{ \{ij\} } h_{ij},\quad h_{ij} = \frac{\vec{S}_i\cdot\vec{S}_j}{(|i-j|+1)^\alpha},
    \label{eq:LSM1d}
\end{equation}
where $\{ ij \}$ includes all pairs of lattice sites. Type II is defined as
\begin{equation}
    H_{II}  = \sum_{ \{ij\} } h_{ij},\quad  h_{ij} = \frac{O_i O_j}{(|i-j|+1)^\alpha},
    \label{eq:LSM1ddensity}
\end{equation}
where $O_i$ and $O_j$ are local operators acting near $i$ and $j$, respectively, and both individually $SO(3)$ symmetric, \eg $O_i=\vec{S}_i\cdot\vec{S}_{i+1}$. After the Jordan-Wigner transformation, $H_I$ becomes fermionic Hamiltonians with long-range particle hopping terms, while $H_{II}$ has only short-range hopping but long-range density-density couplings. Later, we will also discuss generalizations to generic $k$-local Hamiltonians. We are interested in the spectrum of the Hamiltonians, particularly whether a unique symmetric ground state with a nonzero excitation gap $\Delta $ persists in the $L\to\infty$ limit.
We present the following result.
{\theorem In $1d$, (1) The Type I Hamiltonian can not have a unique gapped symmetric ground state for $\alpha>2$; (2) The Type II Hamiltonian can not have a unique gapped symmetric ground state for $\alpha>0$.
\label{thm:1d}
}

\emph{Proof.} We follow the original proof of the LSM theorem \cite{lieb1961two}. The idea is to consider the state after a uniform $2\pi$ flux insertion (twist). We demonstrate that when $\alpha>2$, this twisted state has a vanishing excitation energy relative to $H_I$.   

We define the twist operator implementing the flux insertion as
\begin{equation}
    U = \exp(-i\sum_{j=0}^{L-1} \frac{2\pi}{L}j S_j^z).
\end{equation}
Let the ground state of $H$ be denoted as $| \psi_0 \rangle$. The energy difference $\Delta E$ between the twisted state $U | \psi_0 \rangle$ and $| \psi_0 \rangle$ can be expressed as
\begin{equation}
    \begin{split}
        \langle \psi_0 | U^\dagger H U | \psi_0 \rangle - \langle \psi_0 | H | \psi_0 \rangle & = \langle \psi_0 | U^\dagger [H, U] | \psi_0 \rangle \\
        & \leq \langle \psi_0 | [U^\dagger, [H, U]] | \psi_0 \rangle.
    \end{split}
\end{equation}
We then utilize the explicit expression of $U$ to compute the energy difference, and thus bound the excitation gap as
\begin{equation}
    \begin{split}
        &\Delta E  \leq  2 \sum_{\{ ij \} } \langle \psi_0 | \frac{S_i^x S_j^x + S_i^y S_j^y}{(|i-j|+1)^\alpha}[\cos{\frac{2\pi(i-j)}{L}}-1] | \psi_0 \rangle \\
        & \leq \sum_{ \{ij\} }\frac{2S^2}{(|i-j|+1)^\alpha} (\frac{2\pi(i-j)}{L})^2 
        \leq \max[O(L^{-1}),O(L^{2-\alpha})],
    \end{split}
    \label{eq:1dLSMo}
\end{equation}
where $S$ is the spin quantum number of a single site, and we use $0\leq 1- \cos\theta \leq \frac{\theta^2}{2}$. Here we bound the summation at large distances as
\begin{equation}
    \sum_{j=1}^L \frac{1}{(j+1)^{\alpha-2}} \leq \int_{O(1)}^L dx \frac{1}{x^{\alpha-2}} = \max[O(1),O(L^{3-\alpha})].
\end{equation}
Finally, observe that
\begin{equation}
    T U T^\dagger U^\dagger = e^{-i2\pi S_0^z} e^{i\frac{2\pi}{L}\sum_{j=0}^{L-1}S_j^z},
\end{equation}
where $T$ represents lattice translation, \ie $T S_j^z T^\dagger = S_{j+1}^z$. As $|\psi_0\rangle$ is $SO(3)$ invariant and $e^{-i2\pi S_0^z}=-1$ for $S=\frac{1}{2}$, we deduce that $U|\psi_0 \rangle$ and $|\psi_0\rangle$ have different lattice momentum, and therefore are orthogonal. This, along with a similar argument for $H_{II}$, establishes the proof of Theorem \ref{thm:1d}. \qed

Let us note that our conclusion extends to more general $k$-local Hamiltonians (i.e., each term in the Hamiltonian acts on at most $k$ spins, though these $k$ spins can be far apart), which encompass many physical systems, such as van der Waals interactions, ring-exchange interactions \cite{1999spinexch,2002rindexch}, and multi-body interactions emerging in periodically driven systems \cite{2015driven} and trapped ions \cite{blatt2012quantum,2021timec}, provided that an appropriate condition on convergence is imposed. Consider a $k$-local Hamiltonian
\begin{equation}
    H = \sum_Z H_Z,
\end{equation}
where $Z$ is a subset of sites with cardinality $|Z|\leq k$. Assuming $H$ has a unique symmetric ground state $|\psi_0\rangle$, following the same derivation that leads to Eq. \eqref{eq:1dLSMo}, the excitation gap of $H$ can be bounded by
\begin{equation}
    \begin{split}
        \Delta E & \leq  \langle \psi_0 | [U^\dagger, [H, U]] | \psi_0 \rangle\\
        & \leq \sum_Z \Vert U^\dagger h_Z U + U h_Z U^\dagger -2 h_Z   \Vert.
    \end{split}
\end{equation}
Given the $SO(3)$ invariance of each term, it follows that \cite{tasaki2020physics}
\begin{equation}
    U h_Z U^\dagger = \exp{(-i\frac{2\pi}{L} Q_Z)} h_Z \exp{(i\frac{2\pi}{L} Q_Z)},
\end{equation}
where 
\begin{equation}
    Q_Z = \sum_x (x-x_L) S_x^z,
\end{equation}
with the sum running over all sites on which $h_Z$ acts non-trivially, and $x_L$ is the coordinate of the leftmost site among them (starting from 0). The eigenvalue $\mu$ of $Q_Z$ is bounded by $-k R_Z S \leq \mu \leq k R_Z S$, where $R_Z$ is the largest distance between any two sites in $Z$. The operator $h_Z$ can be decomposed as $h_Z = \sum_\mu h_Z^\mu$, such that $[Q_Z, h_Z^\mu] = \mu h_Z^\mu$. Consequently, we have
\begin{equation}
\begin{split}
    \Delta E & \leq \sum_Z \Vert U^\dagger h_Z U + U h_Z U^\dagger -2 h_Z   \Vert \\
    & \leq \sum_Z \sum_\mu 2[1-\cos(\frac{2\pi}{L}\mu)] \Vert h_Z^\mu \Vert \\
    &\leq \sum_Z \frac{4\pi^2 k^2 R_Z^2 S^2}{L^2} \Vert h_Z \Vert \\
    &\leq \frac{\const}{L} \sum_{Z\ni z} \Vert h_Z \Vert R_Z^2,
\end{split} 
\end{equation}
where we note that $\Vert h_Z^\mu \Vert \leq \Vert h_Z \Vert$, and use translation symmetry to arrive at the last line, where the sum includes all terms that act non-trivially on a single site $z$. Therefore, for a generic $k$-local Hamiltonian, if there exists $\epsilon > 0$ such that
\begin{equation}
    \sum_{Z\ni z} \Vert h_Z \Vert R_Z^2 < L^{1-\epsilon},
    \label{eq:1dconvergence}
\end{equation}
the excitation gap vanishes algebraically for large $L$. The Type I Hamiltonian in Eq.\eqref{eq:LSM1d} with $\alpha>2$ is a special case that satisfies this convergence condition.

On the other hand, for a special subclass of $k$-local Hamiltonians that involve only long-range couplings among $SO(3)$ singlet local operators (or solely ``density-density" couplings after the Jordan-Wigner transformation), the eigenvalue $\mu$ in the commutator $[Q_Z, h_Z^\mu] = \mu h_Z^\mu$ is bounded by a constant that does not depend on the diameter $R_Z$. In this case, if there exists an $\epsilon > 0$ such that
\begin{equation}
    \sum_{Z\ni z} \Vert h_Z \Vert < L^{1-\epsilon},
\end{equation}
then $\Delta E$ would vanish algebraically as $L \to \infty$. The Type II Hamiltonian in Eq.\eqref{eq:LSM1ddensity} with $\alpha > 0$ is an example of such Hamiltonians.

\emph{Higher dimensions.} Now let us consider a $d$-dimensional lattice where one specific dimension, referred to as the $x$ direction, has a size denoted by $L$ (chosen to be even). The Hamiltonian exhibits translational invariance and periodicity along the $x$ direction. All other dimensions are referred to as $y$. The volume of the system, \ie the total number of unit cells, is denoted as $V$. Similar to Ref. \cite{2004Hastings}, it is important to assume that the system's width, defined as $V/L$, is an odd number. This assumption will be used to prove the orthogonality of the twisted state below. The thermodynamic limit is characterized by scaling the sizes of all dimensions to infinity at the same rate, specifically $V\simeq O(L^d)$. The proof in $1d$ does not directly apply, even for finite-range Hamiltonians, as the energy of $U|\psi_0\rangle$ is bounded only by $O(\frac{V}{L}\cdot\frac{1}{L})$. Nonetheless, we still have:

{\theorem
In $d$-dimensions, (1) The Type I Hamiltonian can not have a unique gapped symmetric ground state for $\alpha > \max(3d,4d-2)$. (2) The Type II Hamiltonian can not have a unique gapped symmetric ground state for $\alpha > \max(3d-1,4d-3)$. The theorem holds when $V/L$ is an odd number.
\label{thm:higherd}
}

Before presenting the proof, we have an additional comment. Instead of the specific form in Eq.\eqref{eq:LSM1d} and \eqref{eq:LSM1ddensity}, Theorem \ref{thm:higherd} actually applies to generic $k-$local Hamiltonians, as long as a certain LR bound holds. Consider a Hamiltonian of the form $H = \sum_Z H_Z$, where each term $H_Z$ is supported on a subset of sites denoted by $Z$. As discussed in Ref. \cite{2018Else}, the LR bound holds when
\begin{itemize}
    \item The long-range interaction decays sufficiently fast, \ie for all sites in the entire lattice $\Lambda$,
    \begin{equation}
        \sup_{z\in \Lambda} \sum_{Z\ni z:\mathrm{diam}(Z)\geq R } \Vert H_Z \Vert \leq \frac{\const}{R^{\alpha-d}},
        \label{eq:fastdecay}
    \end{equation}
    with $\alpha>2d$ and $\mathrm{diam}(Z)$ denoting the largest distance between any two points in $Z$;
    \item and
    \begin{equation}
        \sup_{z\in \Lambda} \sum_{y\in \Lambda}\sum_{Z\ni y,z} \Vert H_Z \Vert < \infty,
        \label{eq:extensive}
    \end{equation}
    which ensures that the sum of the norms of all terms acting on any fixed site is finite.
\end{itemize}
Theorem \ref{thm:higherd} applies to generic $k$-local Hamiltonians that meet these two requirements. Specifically, such a Hamiltonian cannot have a unique gapped symmetric ground state when the parameter $\alpha$ in Eq \eqref{eq:fastdecay}, which governs the decay of the interaction, satisfies $\alpha > \max(3d, 4d-2)$ for generic $k$-local interactions, or $\alpha > \max(3d-1, 4d-3)$ for $k$-local interactions among $SO(3)$ singlet local operators. One can verify that Eq. \eqref{eq:LSM1d}
and Eq. \eqref{eq:LSM1ddensity} are examples of such Hamiltonians. The reason for distinguishing between Type I and Type II will become clear when we construct the Hamiltonian with a twist. Lastly, note that the two requirements above on the Hamiltonian differ from the convergence condition \eqref{eq:1dconvergence} specific to $1d$, as the variational state for which the excitation energy is estimated is constructed in a distinct way.

\emph{Proof of Theorem \ref{thm:higherd}.} We follow the approach described in Ref. \cite{2004Hastings}, which is a proof by contradiction. We start by assuming the existence of a unique gapped ground state. Leveraging the LR bound and bounding the norm of long-range couplings, we can demonstrate the insensitivity of the energy to a twisted boundary condition. Namely, we construct a low-lying twisted state with energy algebraically small in $L$. Subsequently, we demonstrate that for spin-$\frac{1}{2}$ per unit cell, the $2\pi$-twisted state has a distinct lattice momentum compared to the initial ground state, and thus must be orthogonal to it. The presence of a low-energy twisted state then leads to a contradiction, thereby establishing the theorem.

\begin{figure}
\begin{center}
  \includegraphics[width=.25\textwidth]{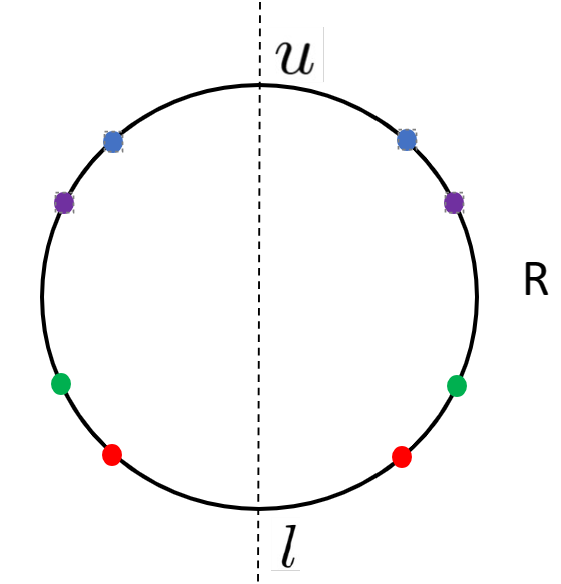} 
\end{center}
\caption{
Plot of the system, showing only the $x$ direction. All sites to the right of the dashed line belong to the right half $R$. The lower quarter $l$ (the shorter interval between the two red dots) contains sites with $x$ coordinates in $[\frac{7L}{8},\frac{L}{8}-1]$. The upper quarter $u$ (the interval between the two blue dots) contains sites with $x\in[\frac{3L}{8},\frac{5L}{8}-1]$. For later convenience, we define the region $l'$ to be sites with $x\in[\frac{13L}{16},\frac{3L}{16}-1]$ (interval between the green dots), and $u'$ to be sites with $x\in[\frac{5L}{16},\frac{11L}{16}-1]$ (interval between the purple dots).
}
\label{Fig:system}
\end{figure}

We focus on a Type I Hamiltonian $H$. To construct a state with a twisted boundary condition, we begin by constructing a twisted Hamiltonian. Unlike Hamiltonians with only finite-range interactions, where a symmetry defect can be introduced by modifying only terms in the vicinity of the twist, constructing a twisted Hamiltonian with long-range interactions is more intricate. Let us introduce a partial spin rotation operator:
\begin{equation}
U_R(\theta) = \exp[-i\theta\sum_{j\in R} S_j^z],
\end{equation}
where the summation extends over all spins on the right half of the lattice, see Fig \ref{Fig:system}. We then denote the Hamiltonian terms supported entirely in the lower (upper) quarter as $H_l$ ($H_u$), \ie $H_l = \sum_{i,j\in l} h_{ij}$. The twisted Hamiltonian is defined as follows:
\begin{equation}
H^{tw}_\theta = H_\theta^l+ H - H_l,
\label{eq:twistedH}
\end{equation}
where $H_\theta^l = U_R^\dagger(\theta) H_l U_R(\theta)$. For later convenience, we also define a Hamiltonian $H_\theta = U_R^\dagger(\theta) H U_R(\theta)$. Compared to $H$, in $H_\theta$, all terms supported simultaneously on the left and right halves are modified. By definition, $H_\theta$ exhibits exactly the same spectrum for all $\theta$, assumed to have a nonzero gap $\Delta$.

\emph{The twisted state.} We introduce the twisted state $|\psi_\theta \rangle$, with its density matrix denoted as $\rho_\theta$, using quasi-adiabatic evolution \cite{2005quasiadiabatic,2004Hastings}. The density matrix of the unique ground state of $H_\theta$ (\ie $|\phi_\theta\rangle:= U^\dagger_R(\theta)|\phi_0\rangle$) is referred to as $\sigma_\theta$. Without loss of generality, we choose $H_\theta\sigma_\theta = E_0 \sigma_\theta =  0$. Define the twisted state $\rho_\theta$ as
\begin{equation}
    \partial_\theta \rho_\theta =  \int_0^{L^\beta} d\tau [A^+(i\tau) - A^-(-i\tau),\rho_\theta],
    \label{eq:twisted}
\end{equation}
with $\rho_0 = \sigma_0$. Here $\beta$ is a control parameter and
\begin{equation}
    A^+(i\tau ) = -\frac{1}{2\pi i} e^{-(\Delta\tau)^2/2q}\int_t \frac{\partial_\theta H_\theta^{tw}(t)}{t+i\tau} e^{-(\Delta t)^2/2q},
    \label{eq:adiabatic}
\end{equation}
where $\partial_\theta H_\theta^{tw}(t) = e^{iH_\theta t} \partial_\theta H_\theta^{tw} e^{-iH_\theta t}$, $A^-(-i\tau) = (A^+(i\tau))^\dagger$ and $q$ is another adjustable parameter. For later convenience, we also define
\begin{equation}
B^+(i\tau ) = -\frac{1}{2\pi i} e^{-(\Delta\tau)^2/2q}\int_t \frac{\partial_\theta H_\theta(t)}{t+i\tau} e^{-(\Delta t)^2/2q},
\label{eq:exactadiabatic}
\end{equation}
where $\partial_\theta H_\theta(t) = e^{iH_\theta t} \partial_\theta H_\theta e^{-iH_\theta t}$ and $B^-(-i\tau) = (B^+(i\tau))^\dagger$. Hereafter, unless otherwise specified, all integration regions are $(-\infty,\infty)$. Note that $\rho_\theta$ is always a pure state for all $\theta$, since commutation with an anti-Hermitian operator generates a unitary rotation. Our goal is to bound the energy of $\rho_\theta$ relative to the Hamiltonian $H_\theta^{tw}$, \ie $\tr(\rho_\theta H_\theta^{tw})$.

To do so, we examine the reduced density matrices in $\Bar{u}$ (the region complementary to the upper quarter), denoted as $\rho_\theta^1$ and $\sigma_\theta^1$, respectively. Specifically, $\rho_\theta^1 = \tr_u(\rho_\theta)$. We aim to demonstrate that
\begin{equation}
    \begin{split}
        \partial_\theta \rho_\theta^1 - \partial_\theta \sigma_\theta^1 
        = e_\theta
    \end{split}
    \label{eq:1error}
\end{equation}
has a small trace norm. Physically, this implies that in the region $\Bar{u}$, the state $\rho_\theta$ closely approximates the state $\sigma_\theta$. We will also show that $\rho_\theta^2 = \tr_{l}(\rho_\theta)$ closely approximates $\sigma_0^2= \tr_{l}(\sigma_0)$, \ie the initial ground state.

To better understand why the error in Eq. \eqref{eq:1error} is small, consider the limit $q \to \infty$ and $\beta\to\infty$, and temporarily substitute $A^{+,-}$ in Eq. \eqref{eq:twisted} with $B^{+,-}$. One can then verify that $\rho_\theta$ precisely matches $\sigma_\theta$, the ground state with two opposite twists at $x=0$ and $x=\frac{L}{2}$, for all $\theta$.\footnote{This can be observed through perturbation theory, \ie $\partial_\theta | \phi_\theta \rangle = \frac{1}{E_0-H_\theta}\partial_\theta H_\theta |\phi_\theta\rangle = -\int_0^\infty d\tau \partial_\theta H_\theta(i\tau)|\phi_\theta\rangle$.} It is found that by choosing $q$ and $L^\beta$ to be finite but large, $\rho_\theta$ can still well approximate $\sigma_\theta$. When the Hamiltonian $H_\theta$ has a gap $\Delta >0$, the LR bound enables us to demonstrate that the impact of introducing the two twists in $\sigma_\theta$ is primarily localized within $l$ and $u$ respectively.\footnote{The reason is that although $H_\theta$ is long-ranged, the main contribution to $\partial_\theta H_\theta$ comes from terms in close proximity to the twists. Consequently, $B^{+,-}$, as defined in Eq. \eqref{eq:exactadiabatic}, is essentially supported within $u$ and $l$ due to the LR bound and the Gaussian that truncates the integral for large $t$.} The motivation behind employing $A^{+,-}$ in Eq. \eqref{eq:twisted} is then straightforward: rather than $\sigma_\theta$ with two opposite twists, we aim to construct a state $\rho_\theta$ with a \emph{single} twist at $x=0$. Consequently, the LR bound suggests that the part of the system within $\overline{u}$ resembles the state $\sigma_\theta$, while $\rho_\theta^2$ remains largely unaltered from $\sigma_0^2$. The reason for considering the reduced density matrices in two overlapping regions $\overline{l}$ and $\overline{u}$ will become apparent when we estimate the energy $\tr(\rho_\theta H_\theta^{tw})$.

We now provide a more precise characterization of the error in Eq \eqref{eq:1error}. The error can be expressed as $e_\theta = e_a + e_b$, where
\begin{equation}
    \begin{split}
        e_a =& \int_0^{L^\beta} d\tau \tr_u\{ [A^+(i\tau)-A^-(-i\tau), \rho_\theta - \sigma_\theta] \}, \\
        e_b =& \int_0^{L^\beta} d\tau \tr_u\{ [A^+(i\tau)-A^-(-i\tau), \sigma_\theta] \} - \partial_\theta \sigma_\theta^1.
    \end{split}
    \label{eq:twoerrors}
\end{equation}
Let us first analyze $e_a$. We observe that in $A^+(i\tau)$, and similarly for $A^-$, the integral over the region $|t|\geq L^\kappa$ contributes with an operator norm upper-bounded by
\begin{equation}
    \begin{split}
      \frac{\Vert \partial_\theta H^{tw}_\theta \Vert}{2\pi} \int_{|t|>L^\kappa} \frac{e^{-(\Delta t )^2/2q}}{t}\leq \frac{\Vert \partial_\theta H_\theta^{tw} \Vert}{2\pi\Delta L^\kappa} \sqrt{\frac{\pi q}{2}} e^{-(\Delta L^\kappa)^2/2q}.
    \end{split}
    \label{eq:ealarget}
\end{equation}
Thus, if we choose $q = L^a$ and $2\kappa>a$, this piece will decay super-polynomially in $L$. After commuting with $\rho_\theta - \sigma_\theta$ in Eq. \eqref{eq:twoerrors}, this piece yields a term with a small trace norm in $e_a$. Consequently, we can disregard the integral over $t$ in the region $|t|\geq L^\kappa$.

We show that, up to a small error, $e_a$ is only a unitary rotation of $\rho_\theta^1 - \sigma_\theta^1$, thereby not increasing its trace norm. To achieve this, we aim to demonstrate that $A^{+,-}$ can be approximated by an operator acting solely within $\overline{u}$, \ie
\begin{equation}
\delta A^+(i\tau) = A^+(i\tau) - \frac{\tr_u[A^+(i\tau)]}{\tr_u \mathbbm{1}_u} \mathbbm{1}_u
\label{eq:choppingoff}
\end{equation}
has a small operator norm. Here, $\mathbbm{1}_u$ represents the identity operator in $u$. The physical interpretation is that each term of $\partial_\theta H_\theta^{tw}$ acts solely within $l$, thus having a minimum distance of $\frac{L}{4}$ from $u$. When $\alpha>2d$, $\partial_\theta H_\theta^{tw}(t)$ can be approximated by an operator within an algebraic light cone \cite{2015Feig,2020hierarchy,2021LRbound,2018Else}. Consequently, within the time range $|t|\leq L^\kappa$ (with $\kappa$ chosen to be sufficiently small), $\delta A^{+,-}$ exhibits a small operator norm. Once we have bounded the operator norm of $\delta A^{+,-}$, $e_a$ simplifies to a unitary rotation of $\rho_\theta^1 - \sigma_\theta^1$, up to an error with a small trace norm, given by:
\begin{equation}
\begin{split}
    e_a = & \int_0^{L^\beta} d\tau  [\frac{\tr_u(A^+(i\tau)-A^-(-i\tau))}{\tr_u \mathbbm{1}_u}, \tr_u(\rho_\theta - \sigma_\theta)] \\
    & + \const \cdot L^{\frac{a}{2}(d+1)+\min(\beta,\frac{a}{2})-(\alpha-2d+1)}.
\end{split}
\label{eq:eaorder}
\end{equation}
We present the detailed calculation of Eqs \eqref{eq:choppingoff} and \eqref{eq:eaorder} in the Supplemental Material (SM) \cite{Supplementary}.

We now turn our attention to $e_b$. We claim that in the first term of $e_b$ (the commutator), substituting $A^{+,-}$ with $B^{+,-}$ introduces only a small error. Let us evaluate the error $\delta$ resulting from this substitution,
\begin{equation}
    \begin{split}
        \delta = & -\frac{1}{2\pi i} \int_0^{L^\beta} d\tau e^{-(\Delta\tau)^2/2q}\\ 
        & \times\int_t \frac{[\partial_\theta H^{tw}_\theta-\partial_\theta H_\theta](t)}{t+i\tau} e^{-(\Delta t)^2/2q}-\hc.
        \end{split}
        \label{eq:errorreplace}
\end{equation}
A crucial observation is that $\partial_\theta H^{tw}_\theta-\partial_\theta H_\theta$, and therefore $\delta$, can be divided into two categories, $\delta = \delta_1+\delta_2$: (1) $\delta_1$ arising from terms entirely within $u$, see Eq \eqref{eq:errordelta1} below; (2) $\delta_2$ from all remaining terms in $\partial_\theta H^{tw}_\theta-\partial_\theta H_\theta$, denoted as $H_m$, constituting a total of $O(L^{2d})$ terms. Notably, each term in $H_m$ acts on two spins separated by a distance of at least $\frac{L}{8}$, thus possessing an operator norm of $O(\frac{1}{L^\alpha})$. The contribution from terms in the second category to $\delta$ is upper bounded by the Shanti's bound \cite{2007multidimensional}
\begin{equation}
\begin{split}
    \Vert \delta_2  \Vert  \leq  \frac{\Vert H_m \Vert}{2} \sqrt{\frac{2\pi q}{\Delta^2}} \leq \const\cdot L^{2d-\alpha+\frac{a}{2}}.
\end{split}
\label{eq:delta2}
\end{equation}
After commuting with $\sigma_\theta$ in Eq. \eqref{eq:twoerrors}, $\delta_2$ contributes a term with a trace norm bounded by the same order as specified in Eq. \eqref{eq:delta2} for $e_b$. 

We now consider $\delta_1$. Define $H_\theta^u = U_R^\dagger(\theta) H_u U_R(\theta)$. One has ($\partial_\theta H_\theta^u(t): = e^{iH_\theta t} \partial_\theta H_\theta^u e^{-iH_\theta t}$)
\begin{equation}
\begin{split}
        \delta_1 = -\frac{1}{2\pi i} \int_0^{L^\beta} d\tau e^{-(\Delta\tau)^2/2q}\int_t \frac{\partial_\theta H_\theta^u (t)}{t+i\tau} e^{-(\Delta t)^2/2q}-\hc.
    \end{split}
    \label{eq:errordelta1}
\end{equation}
In SM \cite{Supplementary}, it is shown that $\tr_u[\delta_1,\sigma_\theta]$ has a trace norm bounded by
\begin{equation}
    \begin{split}
        \Vert \tr_u[\delta_1,\sigma_\theta] \Vert_1 \leq &\max[O(L^{\frac{a}{2}(d+1)+\min(\beta,\frac{a}{2})-(\alpha-2d+1)}),\\
        & O(L^{2d-\alpha+\frac{a}{2}})].
    \end{split}
    \label{eq:finaldelta1}
\end{equation}
Physically, $\delta_1$ introduces a twist at $x=\frac{L}{2}$, whose effect is primarily localized in $u$ due to the LR bound. Combined with $\delta_2$, we conclude that up to small errors (whose trace norms are described by Eqs \eqref{eq:delta2} and \eqref{eq:finaldelta1}, we can substitute $A^{+,-}$ by $B^{+,-}$ in $e_b$.

In SM \cite{Supplementary}, we demonstrate that
\begin{equation}
    \int_0^{L^\beta} d\tau \tr_u\{ [B^+(i\tau)-B^-(-i\tau), \sigma_\theta] \} - \partial_\theta \sigma_\theta^1
    \label{eq:ebfinal}
\end{equation}
has a trace norm that decays super-polynomially in $L$, thereby establishing that $\Vert e_b \Vert_1$ is bounded from above by Eqs \eqref{eq:delta2} and \eqref{eq:finaldelta1}. Physically, in the presence of a nonzero gap, the quasi-adiabatic evolution continues to give a good approximation to $\sigma_\theta$ when $q$ and $L^\beta$ are finite but large, as mentioned earlier, with corrections that are super-polynomially small in $L$.

Using the same reasoning, one can show that up to a small error, the reduced density matrix in $\Bar{l}$, $\rho_\theta^2$, can be approximated by $\sigma_0^2$. Namely, we have $\partial_\theta \rho_\theta^2 =e_\theta'$ with
\begin{equation}
    \begin{split}
         \Vert e_\theta' \Vert_1 \leq & \const\cdot \max[L^{\frac{a}{2}(d+1)+\min(\beta,\frac{a}{2})-(\alpha-2d+1)},\\
         & L^{2d-\alpha+\frac{a}{2}}].
    \end{split}
\end{equation}
In fact, since $\partial_\theta H_\theta^{tw}$ is entirely within $l$, $e_\theta'$ only receives contributions similar to those described in Eq \eqref{eq:finaldelta1}.

We can now determine the energy of the twisted state $\rho_\theta$ with respect to $H_\theta^{tw}$. 
The first term in Eq \eqref{eq:twistedH} resides entirely within $\Bar{u}$, leading to $\tr[ H_l^\theta \rho_\theta] \simeq \tr_{\Bar{u}}[U_R^\dagger(\theta) H_l U_R(\theta)\sigma_\theta^1] = \tr(H_l\sigma_0)$ up to an error of order $O(\Vert H_l \Vert\cdot \Vert e_\theta \Vert_1)$. The second part of $H_\theta^{tw}$, $H-H_l$, can be classified into three categories: (1) For $H_{\Bar{l}}$, \ie terms entirely supported in $\Bar{l}$, we obtain $\tr[H_{\Bar{l}}\rho_\theta] \simeq \tr(H_{\Bar{l}}\sigma_0)$ up to an error bounded by $O( \Vert H_{\Bar{l}} \Vert \cdot \Vert e_\theta' \Vert_1)$. (2) $H_{\Bar{u}}^L$ and $H_{\Bar{u}}^R$, supported in $\Bar{u}$ and acting only in the left/right half, yield $\tr[(H_{\Bar{u}}^L+H_{\Bar{u}}^R)\rho_\theta] \simeq \tr[(H_{\Bar{u}}^L+H_{\Bar{u}}^R)\sigma_0]$ up to an error bounded by $O[ (\Vert H_{\Bar{u}}^L \Vert + \Vert H_{\Bar{u}}^R \Vert) \cdot \Vert e_\theta \Vert_1]$. (3) Finally, $K = H-H_l-H_{\Bar{l}}-H_{\Bar{u}}^L-H_{\Bar{u}}^R$ contains $O(L^{2d})$ terms, each acting on two spins with a distance of at least $\frac{L}{8}$. Consequently, $\Vert K \Vert \leq \const\cdot L^{2d-\alpha}$. (We analyze reduced density matrices in two overlapping regions $\overline{l}$ and $\overline{u}$, precisely to suppress the contribution from terms acting simultaneously on $u$ and $l$.) Combining all results, for $\alpha>2d$ we have
\begin{equation}
    \tr(H_\theta^{tw} \rho_\theta) \leq \const \cdot L^d ( \Vert e_\theta \Vert_1 + \Vert e_\theta' \Vert_1 ) +O(L^{2d-\alpha}),
\end{equation}
where we utilized the fact that for $\alpha>d$, $\Vert H \Vert \simeq O(L^d)$. It is evident that for any $\alpha>3d$, one can always choose $a$ and $\beta$ sufficiently small such that the energy of $\rho_\theta$ (particularly, $\rho_{2\pi}$) vanishes algebraically with $L$. This completes the construction of a low-lying twisted state.

\emph{Lattice momentum.} Finally, we demonstrate that the twisted state $\rho_{2\pi}$ has a momentum distinct from the ground state $\sigma_0$, thereby being orthogonal to it. We define the translation in the presence of two twists (at links along the $x$ direction with coordinates $(L-1,0)$ and $(\frac{L}{2}-1,\frac{L}{2})$, respectively) as
\begin{equation}
    T_{\theta,\theta'} = \exp[i\theta S_z^{(0)} + i\theta' S_z^{(\frac{L}{2})}]T,
\end{equation}
where $T$ represents the ordinary lattice translation along the $x$ direction, $S_z^{(0)} = \sum_y S^z_{(0,y)}$ denotes the summation of all spin-$z$ operators on sites with $x$ coordinate 0, and similarly for $S_z^{(\frac{L}{2})}$. It can be verified that (1) $T_{\theta,-\theta}$ is a symmetry of $H_\theta$; (2) the state $\phi_\theta$ is an eigenstate of $T_{\theta,-\theta}$ with $T_{\theta,-\theta}\phi_\theta = \phi_\theta$. Here without loss of generality, we assume that $T\phi_0 = \phi_0$.

We now demonstrate that $\partial_\theta (T_{\theta,0}\psi_\theta-\psi_\theta)$ has a small norm for $0\leq \theta \leq 2\pi$, where $\psi_\theta$ is the twisted state, \ie the state vector for $\rho_\theta$. This would imply that $\psi_{2\pi}$ is almost an eigenvector of $T_{2\pi,0}$ with an eigenvalue of 1. As the width of the system is odd, we have $T = -T_{2\pi,0}$. Hence, $\psi_{2\pi}$ exhibits opposite momentum compared to $\psi_0 = \phi_0$. The derivative can be computed as
\begin{equation}
    \begin{split}
        &\partial_\theta (T_{\theta,0}\psi_\theta-\psi_\theta) = O_1  \psi_\theta + \Tilde{O}_1 (T_{\theta,0}\psi_\theta-\psi_\theta), \\
        & O_1 = i S_z^{(0)} + \int_0^{L^\beta} d\tau [T_{\theta,0}A^+(i\tau) T_{\theta,0}^\dagger - A^+(i\tau) - \hc],
    \end{split}
    \label{eq:momentumd}
\end{equation}
where $\Tilde{O}_1$ is anti-Hermitian, inducing only a unitary rotation without altering the norm of $T_{\theta,0}\psi_\theta-\psi_\theta$. We then bound the square norm of the first term in Eq \eqref{eq:momentumd}, specifically $\tr(O_1^\dagger O_1 \rho_\theta)$.

We provide an overview of the argument here, leading to the conclusion that for $\alpha>4d-2$, $\psi_{2\pi}$ has lattice momentum $\pi$, while reserving a detailed discussion for SM \cite{Supplementary}. Using a reasoning similar to that leading to Eq \eqref{eq:eaorder}, the operator $O_1$ can be approximated by $O_{l'}$, an operator supported solely in $l'$ (see Fig. \ref{Fig:system}), with an error bounded by the second line of Eq \eqref{eq:eaorder}. Here, $O_{l'}$ is defined by truncating the effect of $O_1$ in $\Bar{l'}$, akin to Eq \eqref{eq:choppingoff}. Consequently, we can approximate $\tr(O_1^\dagger O_1 \rho_\theta)$ by $\tr(O_{l'}^\dagger O_{l'} \rho_\theta)$, further approximated by $\tr(O_{l'}^\dagger O_{l'} \sigma_\theta)$ since $l'\subset \Bar{u}$. Finally, we claim that $\tr(O_{l'}^\dagger O_{l'} \sigma_\theta)$ vanishes in the $L\to \infty$ limit. To demonstrate this, observe that $O \phi_\theta :=\partial_\theta (T_{\theta,-\theta} \phi_\theta - \phi_\theta) =  0$. Additionally, $O$ can be approximated by $O \simeq O_{l'} + O_{u'}$, where $O_{u'}$ is an operator akin to $O_{l'}$, but supported entirely within $u'$. Hence, we have $0=\tr(O^\dagger O \sigma_\theta) = 2\tr(O_{l'}^\dagger O_{l'}\sigma_\theta) + \tr(O_{l'}^\dagger O_{u'} \sigma_\theta) + \tr(O_{u'}^\dagger O_{l'} \sigma_\theta)$. It has been established that if $\sigma_\theta$ is a unique gapped ground state of a power-law Hamiltonian, the correlation function $\tr(O_{l'}^\dagger O_{u'} \sigma_\theta) + \tr(O_{u'}^\dagger O_{l'} \sigma_\theta)$ between two operators $O_{l'}$ and $O_{u'}$ (separated by at least a distance $\frac{L}{8}$) decays algebraically in $L$ \cite{2020hierarchy}. This ultimately bounds the norm of $O_1 \psi_\theta$, establishing that $\psi_{2\pi}$ possesses momentum $\pi$. This completes the proof of Theorem \ref{thm:higherd}. \qed

We can establish the second part of Theorem \ref{thm:higherd} following the same line of reasoning. For Type II Hamiltonians (or their generic $k$-local generalizations, i.e., Hamiltonians that involve $k$-body couplings of $SO(3)$ singlet local operators), the same LR bound still applies when the two conditions \eqref{eq:fastdecay} and \eqref{eq:extensive} are satisfied. The only distinction lies in the introduction of a twist: in the construction of $H_\theta^{l}$, $H_\theta$ and $H_\theta^u$, only the terms acting non-trivially on the twist are modified. For the simplest example, where $H = \sum_{{ij}} h_{ij}$ with $h_{ij} = \frac{O_i O_j}{(|i-j|+1)^\alpha}$, the twist only affects terms where either $O_i$ or $O_j$ acts precisely on the twist. In contrast, for a Type I Hamiltonian, in $H_\theta^l$ in Eq.\eqref{eq:twistedH}, all terms straddling between the left and right halves need to be twisted. This distinction results in smaller errors for Type II cases, thereby extending the applicability of the Theorem. We present a detailed discussion in SM \cite{Supplementary}. 

\emph{Summary and outlook.} In this study, we have established an LSM theorem applicable to systems with long-range interactions. The proof involves the construction of a low-energy twisted state through quasi-adiabatic evolution and the LR bound for power-law Hamiltonians, which is then shown to be orthogonal to the ground state when each unit cell has spin-$\frac{1}{2}$. Moreover, in $1d$, we enhance the theorem's applicability by explicitly constructing a low-lying state featuring a uniform $2\pi$ twist.

The result we obtained is weaker than that in Ref. \cite{2004Hastings}. In that work, the excitation gap is bounded by $\Delta\leq\const\frac{\ln L}{L}$, whereas here we only conclude that a nonzero gap $O(L^0)$ is forbidden. It is an interesting future direction to explore a similar finer bound on $\Delta $ for power-law interacting Hamiltonians, \ie understanding how quickly $\Delta $ vanishes at large $L$. Moreover, extending the proof to other (\eg discrete) onsite symmetries, with a projective representation per site, remains an open question. Another interesting question is whether the bounds given in this work are sharp, \ie can we find counterexamples where the theorems do not hold when the bounds on the decaying power are violated.

\begin{acknowledgements}

I am grateful to Meng Cheng, Tarun Grover, Chong Wang, Weicheng Ye, and especially Michael Levin for discussions. I also thank Yushao Chen for help with Mathematica. I am supported in part by Simons Investigator Award number 990660 and by the Simons Collaboration on Ultra-Quantum Matter, which is a grant from the Simons Foundation (No. 651440).

\emph{Note added}: Upon completion of the manuscript, I became aware of another work that explores
LSM with long-range interactions \cite{LSMZou2024},
which would appear on arXiv on the same day.

\end{acknowledgements}

\bibliography{Ref.bib}
\clearpage

\appendix

In this Supplemental Material, we provide technical details regarding several equations in the main text.

\emph{Eqs \eqref{eq:choppingoff} and \eqref{eq:eaorder}}.

In the definition of $e_a$ in Eq \eqref{eq:twoerrors}, $A^{+,-}$ involves an integral over $t$. We have shown in Eq \eqref{eq:ealarget} that the contribution from $|t|\geq L^\kappa$ is negligible. Now we argue that $A^{+,-}$ can be approximated by an operator acting only in $\Bar{u}$, namely Eq \eqref{eq:choppingoff} has a small operator norm.

We have the following expression:
\begin{equation}
    \frac{\tr_u[A^+(i\tau)]}{\tr_u \mathbbm{1}_u} \mathbbm{1}_u = \int d\mu(U) U A^+(i\tau) U^\dagger, 
\end{equation}
where $U$ is the unitary group acting on $u$, and $\mu(U)$ is the Haar measure for $U$ \cite{2006LRbound}. As a result, 
\begin{equation}
   \Vert \delta A^+(i\tau) \Vert \leq \int d\mu(U) \Vert [U, A^+(i\tau)] \Vert.
\end{equation}
To bound the RHS, we compute the norm of the commutator $\Vert [\partial_\theta H^{tw}_\theta(t),U] \Vert$, where $U$ is an operator supported entirely in the upper quarter with unit norm. A crucial observation is that $\partial_\theta H_\theta^{tw}$ contains only terms in the lower quarter $l$, and supported simultaneously in the left and right halves. The minimum distance from each of these terms to the upper quarter is $\frac{L}{4}$. We then apply the LR bound to estimate the commutator as
\begin{equation}
    \begin{split}
        \Vert [\partial_\theta H^{tw}_\theta(t), & U] \Vert \leq \sum_{ij} \Vert [\partial_\theta h_{ij}^\theta(t),U] \Vert \\
        & \leq \const \sum_{ij} \frac{C(t,r_{ij}^u)}{(|i-j|+1)^\alpha}
    \end{split}
    \label{eq:linfluu}
\end{equation}
where (1) $h_{ij}^\theta = U_R^\dagger(\theta)h_{ij} U_R(\theta)$; (2) the pairs $ij$ are such that $i,j\in l$ and one of them is in the left half; (3) $C(t,r)$ is the LR bound,
\begin{equation}
    C(t,r) = \sup_{\Vert A \Vert=\Vert B \Vert=1} \Vert [A(t),B] \Vert,
\end{equation}
with $r$ representing the minimal distance of $A$ and $B$. (4) $r_{ij}^u$ denotes the minimal distance from the pair $ij$ to the upper quarter $u$.

Let us now estimate the magnitude of Eq. \eqref{eq:linfluu}. Firstly, as previously discussed, $r_{ij}^u$ is at least $\frac{L}{4}$. It has been demonstrated that in the presence of power law interactions, when $\alpha>2d+1$, the system still exhibits a linear light cone \cite{2020hierarchy},
\begin{equation}
    C(t,r)\leq \const \cdot |A| \frac{t^{d+1}\log^{2d}r}{(r-vt)^{\alpha-d}},
    \label{eq:LRboundform}
\end{equation}
where $|A|$ represents the number of sites acted upon by $A$, and $v$ is a finite velocity.\footnote{The discussion also applies to $\alpha\in (2d,2d+1]$, where in this range, a nonlinear LR light cone is present \cite{2021LRbound}, given by
\begin{equation}
C(t,r)\leq \const \cdot \frac{t^{(\alpha-d)/(\alpha-2d)}}{r^{\alpha-d}},
\end{equation}
for $t\leq \const\cdot r^{\alpha-2d}$.} $C(t,r)$ is an algebraic function of $t$ and $r$, decaying as $r$ increases. Thus, we can replace $r_{ij}^u$ with $\frac{L}{4}$ in Eq \eqref{eq:linfluu} without altering its order. Secondly, a direct computation indicates that for $\alpha>d+1$, the summation in Eq. \eqref{eq:linfluu} is dominated by pairs $ij$ where $i$ and $j$ are in close proximity, and therefore in the vicinity of the twist. In other words, the summation is on the order of
\begin{equation}
    \sum_{\substack{i\in l\cap L \\ j\in l\cap R}}\frac{1}{(|i-j|+1)^\alpha} \simeq O(L^{d-1}),
    \label{eq:areatwist}
\end{equation}
which corresponds to the area of the twist.

Combining the discussion from the previous paragraph, we can bound the commutator as follows:
\begin{equation}
    \begin{split}
    &\Vert [\int_0^{L^\beta} d\tau \delta A^+(i\tau),U] \Vert\\
     \leq &   \const \int_0^{L^\beta} d\tau e^{-(\Delta\tau)^2/2q} \int_{|t|<L^\kappa} dt \frac{t^{d+1} e^{-(\Delta t)^2/2q}}{ L^{\alpha-2d+1}|t+i\tau|}.
    \end{split}
    \label{eq:deltaA+}
\end{equation}
Here we have chosen $\kappa<1$, which will not lead to any contradictions below, and ignored the logarithmic correction. Two observations are useful for bounding the integral in Eq \eqref{eq:deltaA+}: (1) For $L^\kappa \gg q^\frac{1}{2}= L^\frac{a}{2}$ (\ie $\kappa>\frac{a}{2}$), the integration over $t$ is dominated by $t\simeq O(q^\frac{1}{2}/\Delta)$, therefore the right-hand side of Eq \eqref{eq:deltaA+} has the same order in $L$ as $ \int_0^{L^\beta}d\tau e^{-(\Delta\tau)^2/2q} \frac{q^\frac{d+2}{2}}{L^{\alpha-2d+1}|(q^\frac{1}{2}/\Delta)+i\tau|}$; (2) Doing the integration over $\tau$ leads to the final result,
\begin{equation}
    \Vert [\int_0^{L^\beta} d\tau \delta A^+(i\tau),U] \Vert\leq \const
    \frac{L^\frac{a(d+1)}{2} L^{\min(\beta,\frac{a}{2})}}{ L^{\alpha-2d+1}}.
    \label{eq:eaerror}
\end{equation}
where the constant factor does not scale with $L$.\footnote{Explicitly integrating the expression in Eq. \eqref{eq:deltaA+} results in a special function which shares the same order in $L$ as the expression in Eq. \eqref{eq:eaerror}, accompanied by a Gamma function coefficient.} The same analysis can be applied to $A^-$. Upon commuting with $\rho_\theta-\sigma_\theta$, we deduce that, up to an error whose trace norm is bounded by Eq \eqref{eq:eaerror}, $e_a$ is simply a unitary rotation of $\rho_\theta^1 - \sigma_\theta^1$ and thus does not increase its trace norm. Namely,
\begin{equation}
\begin{split}
    e_a = & \int_0^{L^\beta} d\tau  [\tr_u(A^+(i\tau)-A^-(-i\tau)), \tr_u(\rho_\theta - \sigma_\theta)] \\
    & + \const \cdot L^{\frac{a}{2}(d+1)+\min(\beta,\frac{a}{2})-(\alpha-2d+1)}.
    \label{eq:eaerrorA}
\end{split}
\end{equation}
This is precisely Eq \eqref{eq:eaorder}.

Finally, we note that the analysis above actually applies to generic $k-$local Hamiltonians satisfying Eq. \eqref{eq:fastdecay} and \eqref{eq:extensive}. To demonstrate this, we require an LR bound for generic Hamiltonians satisfying the two properties mentioned above \cite{2018Else},
\begin{equation}
    C(t,r)\leq 2|A|e^{vt-r^{1-\sigma}} +\const\frac{f(vt)}{r^{\sigma\alpha}},
    \label{elseLR}
\end{equation}
where $\sigma$ can take any value within $(\frac{d+1}{\alpha+1},1)$, and $f(vt)$ is upper bounded by 
\begin{equation}
    f(x)\leq \const (x+x^{1+\frac{d}{1-\sigma}}) |A|^{n^*+2},
\end{equation}
with $n^*$ being the smallest integer greater than or equal to $\frac{\sigma d}{\sigma\alpha-d}$. For $k-$local Hamiltonians, $|A|$ is bounded by a constant and is therefore not relevant for the discussion below. We conjecture that all the derivations below, and consequently Theorem \ref{thm:higherd}, also apply to Hamiltonians where the number of spins acted by each term is unbounded, provided, for instance, that the norm decays superpolynomially with the number of spins it acts on. A thorough investigation of more general Hamiltonians is left for future work.

Furthermore, one can demonstrate that Eq.\eqref{eq:areatwist} still holds for generic Hamiltonians satisfying Eq.\eqref{eq:fastdecay}, \ie $\Vert\partial_\theta H_\theta^{tw} \Vert$ is bounded by $O(L^{d-1})$. Repeating the analysis leading to Eq \eqref{eq:eaerror} and \eqref{eq:eaerrorA}, one finds that for generic $k-$local Hamiltonians satisfying Eq \eqref{eq:fastdecay} and \eqref{eq:extensive}, we have 
\begin{equation}
\begin{split}
    &\Vert [\int_0^{L^\beta} d\tau \delta A^+(i\tau),U] \Vert \\
    &\leq \const
    \frac{L^{\frac{a}{2}(\frac{d}{1-\sigma}+1)} L^{\min(\beta,\frac{a}{2})+(1-\sigma)(\alpha-d)}}{ L^{\alpha-2d+1}}.
\end{split}
\label{klocalerrorea}
\end{equation}
and
\begin{equation}
\begin{split}
    e_a = & \int_0^{L^\beta} d\tau  [\tr_u(A^+(i\tau)-A^-(-i\tau)), \tr_u(\rho_\theta - \sigma_\theta)] \\
    & + \const \cdot L^{\frac{a}{2}(\frac{d}{1-\sigma}+1)+\min(\beta,\frac{a}{2})+(1-\sigma)(\alpha-d)-(\alpha-2d+1)}.
\end{split}
\label{eq:klocalea}
\end{equation}
As we will see below, although considering generic $k-$local Hamiltonians complicates the calculation, it does not alter the final results: the essential point is that for any $\alpha>2d-1$, one can always find $\sigma$ arbitrarily close to 1 and $a,\,\beta$ sufficiently small, such that the error in Eq \eqref{klocalerrorea} vanishes in the $L\to \infty$ limit.

\emph{Eq \eqref{eq:finaldelta1}.}

\begin{figure}
\begin{center}
  \includegraphics[width=.25\textwidth]{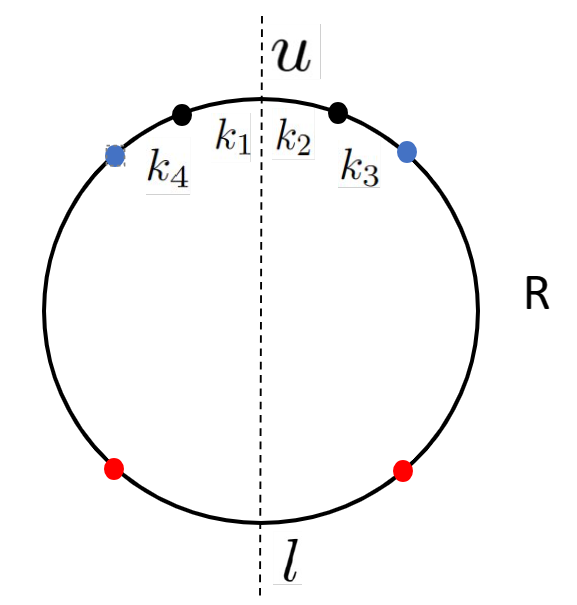} 
\end{center}
\caption{
The region $k_1$ includes sites with $x\in[\frac{L}{2},\frac{9L}{16}-1]$. Similarly, $k_2$ consists of sites with $x\in[\frac{7L}{16},\frac{L}{2}-1]$; $k_3$ comprises sites with $x\in[\frac{3L}{8},\frac{7L}{16}-1]$; and $k_4$ covers sites with $x\in[\frac{9L}{16},\frac{5L}{8}-1]$.
}
\label{Fig:system1}
\end{figure}

Let us now consider the error $\delta_1$ defined in Eq \eqref{eq:errordelta1}. In particular, we demonstrate Eq \eqref{eq:finaldelta1}, \ie $\tr_u[\delta_1,\sigma_\theta]$ has a small trace norm. To show this, consider the trace of this operator with an operator $O$ with unit norm. $O$ must be supported within $\Bar{u}$. One has
\begin{equation}
    \tr_{\Bar{u}}(O\tr_u[\delta_1,\sigma_\theta]) = \tr([O,\delta_1]\sigma_\theta).
    \label{eq:delta1bound}
\end{equation}
We now bound the operator norm of $[O,\delta_1]$. Note that $\partial_\theta H_\theta^u$ can be further divided into two parts, $\partial_\theta H_\theta^u = K_1 + K_2$: (1) The terms $K_1$ that act on two spins in the region $k_1$ and $k_2$ (as depicted in Fig \ref{Fig:system1}) respectively. Note that the minimal distance from each of these terms to $O$ is $\frac{L}{16}$. Following exactly the same argument leading to Eq \eqref{eq:eaerror}, the LR bound constrains the commutator
\begin{equation}
   \begin{split}
       &\Vert [-\frac{1}{2\pi i} \int_0^{L^\beta} d\tau e^{-(\Delta\tau)^2/2q}\int_t \frac{K_1(t)}{t+i\tau} e^{-(\Delta t)^2/2q}-\hc,O] \Vert \\
       & \leq\const\cdot L^{\frac{a}{2}(d+1)+\min(\beta,\frac{a}{2})-(\alpha-2d+1)}.
   \end{split}
   \label{eq:delta1K1}
\end{equation}
For a generic $k-$local Hamiltonian with properties \eqref{eq:fastdecay} and \eqref{eq:extensive}, this commutator is similarly bounded by Eq. \eqref{klocalerrorea}.

(2) $K_2$ includes all other terms in $K$, namely terms that act on either the region $(k_1,k_3)$, $(k_2,k_4)$, or $(k_3,k_4)$. The norm of each term in $K_2$ is upper-bounded by $(\frac{16}{L})^\alpha$, and $K_2$ contains $3\times(\frac{L^d}{16})^2$ terms in total. As a result, we have (Shanti's bound)
\begin{equation}
    \begin{split}
       &\Vert [-\frac{1}{2\pi i} \int_0^{L^\beta} d\tau e^{-(\Delta\tau)^2/2q}\int_t \frac{K_2(t)}{t+i\tau} e^{-(\Delta t)^2/2q}-\hc,O] \Vert \\
       & \leq\const\cdot L^{2d-\alpha+\frac{a}{2}}.
   \end{split}
   \label{eq:delta1K2}
\end{equation}
For general $k-$local Hamiltonians, the property \eqref{eq:fastdecay} ensures that the same order holds. Combining Eqs \eqref{eq:delta1K1} and \eqref{eq:delta1K2}, we arrive at Eq \eqref{eq:finaldelta1}. This is because we have bounded the trace in Eq \eqref{eq:delta1bound} with any arbitrary $O$ with unit norm, thereby bounding the trace norm of $\tr_u[\delta_1,\sigma_\theta]$.

\emph{Eq \eqref{eq:ebfinal}.}

We now demonstrate that the expression in Eq \eqref{eq:ebfinal} has a trace norm that decays super-polynomially in $L$. To see this, let us examine the matrix element of the operator $B^+(i\tau)$,
\begin{equation}
    \begin{split}
         [B^+(i\tau )]_{ab}
        = & \sqrt{\frac{q}{2\pi\Delta^2}} (\partial_\theta H_\theta)_{ab} \int_\omega \Theta(\omega-(E_a-E_b)) \\
        & \times  e^{(E_a-E_b)\tau-(\sqrt{\frac{q}{2\Delta^2}}\omega+\sqrt{\frac{\Delta^2}{2q}}\tau)^2},
    \end{split}
    \label{eq:Bspectrum}
\end{equation}
where $a,b$ label eigenstates of $H_\theta$, and we have used the convolution theorem. When acting on the gapped ground state $\sigma_\theta$, we only need to consider elements with $|E_a-E_b|\geq \Delta$. For sufficiently large $q$ ($q = L^a \gg L^\beta$, \ie $a>\beta$), when (1) $E_a<E_b$, the matrix element $[B^+(i\tau)]_{ab}$ becomes $e^{(E_a-E_b)\tau}[\partial_\theta H_\theta]_{ab}$; (2) for $E_a>E_b$, $B^+(i\tau)$ goes to zero, both with corrections of order $O(e^{-\frac{q}{2\Delta^2}(E_a-E_b)^2})$, decaying super-polynomially in $L$. The same argument applies to $B^-$. Thus, in the presence of a nonzero gap $\Delta$, up to super-polynomial corrections, the first term in Eq \eqref{eq:ebfinal} can be approximated as
\begin{equation}
    -\int_0^{L^\beta} d\tau \tr_u[\partial_\theta H_\theta(i\tau)\sigma_\theta+\sigma_\theta\partial_\theta H_\theta(-i\tau)].
    \label{eq:Badiabatic}
\end{equation}
Subsequently, Eq \eqref{eq:ebfinal} transforms into an integral over the region $\tau>L^\beta$, whose trace norm once again vanishes super-polynomially in the presence of a nonzero $\Delta$. This is evident from a direct calculation, \ie Eq \eqref{eq:ebfinal} becomes
\begin{equation}
    \begin{split}
        -\sum_{a\neq 0}\tr_u  [\sigma_\theta\frac{\partial_\theta H_\theta|a\rangle\langle a|}{E_0- E_a}e^{L^\beta (E_0-E_a)}+\hc],
    \end{split}
\end{equation}
whose trace norm decays super-polynomially in $L$ when $\beta>0$.

\emph{Lattice momentum.}

We now bound the norm square of the first term in Eq \eqref{eq:momentumd}, namely $\tr(O_1^\dagger O_1 \rho_\theta)$. If this quantity vanishes in the $L\to \infty$ limit, we will have $\tr(T_{0,0}\rho_0) = \tr(T_{2\pi,0}\rho_{2\pi}) = -\tr(T_{0,0}\rho_{2\pi})$, indicating that the twisted state $\psi_{2\pi}$ is orthogonal to the original ground state $\phi_0$.

As discussed in Eq. \eqref{eq:choppingoff}, the operator $\int_0^{L^\beta} A^+(i\tau)$ (and thus the translated version $T_{\theta,0}\int_0^{L^\beta} A^+(i\tau) T_{\theta,0}^\dagger$) can be approximated by an operator supported solely in $\Bar{u}$, with an error bounded by Eq \eqref{eq:eaerror}. In fact, this argument allows us to approximate it even further as an operator acting in a smaller region $l'$, with an error of the same order. Therefore, we can express $O_1$ as $O_{l'}+\delta O_1$, where $O_{l'}$ is entirely within $l'$. Consequently, we approximate $\tr(O_1^\dagger O_1 \rho_\theta)$ by $\tr(O_{l'}^\dagger O_{l'} \rho_\theta)$, with an error bound given by:
\begin{equation}
\begin{split}
    &|\tr [O_1 \delta O_1 \rho_\theta]| \leq \Vert O_1 \Vert \cdot \Vert \delta O_1 \Vert \\
    &\leq \const \cdot L^{d-1+\frac{a}{2}}  \cdot L^{\frac{a}{2}(d+1)+\min(\beta,\frac{a}{2})-(\alpha-2d+1)}.
\end{split}
\label{eq:operatorrestriction}
\end{equation}
Here, we utilize: (1) the fact that for $\alpha>d+1$, $\Vert\int_0^{L^\beta} d\tau A(i\tau) -\hc\Vert \leq \const \cdot L^{d-1+\frac{a}{2}}$, as $\Vert \partial_\theta H_\theta^{tw} \Vert\simeq O(L^{d-1})$ is dominated by terms near the twist (see Eq \eqref{eq:areatwist}); and (2) $\Vert \delta O_1 \Vert$ is bounded by Eq \eqref{eq:eaerror}. For $\alpha> 3d-2$, we can choose $a$ and $\beta$ sufficiently small such that this error vanishes in the $L\to\infty$ limit. Similarly, for general $k-$local Hamiltonians, due to Eq \eqref{klocalerrorea}, we have
\begin{equation}
\begin{split}
    &|\tr [O_1 \delta O_1 \rho_\theta]|\\
    &\leq \const \cdot L^{d-1+\frac{a}{2}}  \cdot L^{\frac{a}{2}(\frac{d}{1-\sigma}+1)+\min(\beta,\frac{a}{2})+(1-\sigma)(\alpha-d)-(\alpha-2d+1)}.
\end{split}
\label{eq:operatorrestrictionklocal}
\end{equation}
Again for $\alpha> 3d-2$, we can choose $\sigma$ sufficiently close to 1, and $a$ and $\beta$ sufficiently small, such that this error vanishes in the $L\to\infty$ limit.

Additionally, we have $\tr(O_{l'}^\dagger O_{l'} \rho_\theta) = \tr_{\Bar{u}}(O_{l'}^\dagger O_{l'} \rho^1_\theta) = \tr_{\Bar{u}}(O_{l'}^\dagger O_{l'} \sigma^1_\theta)$, with an error bounded by (H$\mathrm{\ddot{o}}$lder's inequality):
\begin{equation}
\begin{split}
    2\pi \Vert O_{l'} \Vert^2 \cdot \Vert e_\theta \Vert_1 \leq & \const \Vert O_1 \Vert^2 \cdot \Vert e_\theta \Vert_1 \\
    \leq & \const\cdot L^{4d-2-\alpha+\frac{3a}{2}}.
\end{split} 
\label{eq:Holder}
\end{equation}
For $\alpha>4d-2$, we can always choose $a$ and $\beta$ such that this error vanishes as $L$ grows large.

We then claim that $\tr_{\Bar{u}}(O_{l'}^\dagger O_{l'} \sigma^1_\theta) = \tr(O_{l'}^\dagger O_{l'} \sigma_\theta) $ vanishes in the thermodynamic limit. To see this, note that $\partial_\theta (T_{\theta,-\theta} \phi_\theta - \phi_\theta) = O \phi_\theta = 0$, where
\begin{equation}
\begin{split}
    O =& i(S_z^{(0)}-S_z^{(\frac{L}{2})})\\-&\int_0^\infty d\tau [T_{\theta,-\theta}\partial_\theta H_\theta(i\tau) T_{\theta,-\theta}^\dagger -\partial_\theta H_\theta(i\tau) ].
\end{split}
\label{eq:momentum2}
\end{equation}
When $H_\theta$ has a unique gapped ground state, within an error with an operator norm that decays super-polynomially in $L$, we can approximate the second line in Eq \eqref{eq:momentum2} as follows:
\begin{equation}
\begin{split}
\int_0^{L^\beta} d\tau &[T_{\theta,-\theta} (B^+(i\tau)-B^-(-i\tau))T_{\theta,-\theta}^\dagger\\
&-(B^+(i\tau)-B^-(-i\tau))]. 
\end{split}
\label{eq:momentum3}
\end{equation}
This approximation is justified by the same reasoning elucidated below Eq \eqref{eq:Bspectrum}. Furthermore, up to a small error, $B^{+,-}$ in Eq \eqref{eq:momentum3} can be approximated by $\Tilde{B}^{+,-}$, which is defined as
\begin{equation}
    \Tilde{B}^+(i\tau ) = -\frac{1}{2\pi i} e^{-(\Delta\tau)^2/2q}\int_t \frac{\partial_\theta \Tilde{H}_\theta(t)}{t+i\tau} e^{-(\Delta t)^2/2q}.
\end{equation}
where $\Tilde{H}_\theta := H^l_\theta+H^u_\theta$. This approximation is valid because each term in $\partial_\theta H_\theta - \partial_\theta \Tilde{H}_\theta$ operates on two spins separated by a distance of at least $\frac{L}{8}$, and therefore has a small operator norm (the error resulting from this substitution in Eq \eqref{eq:momentum3} is of the same order as that in Eq \eqref{eq:delta2}). Consequently, we can express $O$ as the sum of $O_1'$, $O_2'$, and $\delta O$, with $\Vert \delta O \Vert$ bounded by Eq \eqref{eq:delta2}. Here, $O_1'$ is approximately the operator $O_1$ defined in Eq \eqref{eq:momentumd}, but with $T_{\theta,0}$ replaced by $T_{\theta,-\theta}$:
\begin{equation}
    O_1' = i S_z^{(0)} + \int_0^{L^\beta} d\tau [T_{\theta,-\theta}A^+(i\tau) T_{\theta,-\theta}^\dagger - A^+(i\tau) - \hc].
\end{equation}
Additionally,
\begin{equation}
    O_2' = -i S_z^{(\frac{L}{2})}+ [T_{\theta,-\theta}\delta_1 T_{\theta,-\theta}^\dagger - \delta_1],
    \label{eq:O2'}
\end{equation}
where $\delta_1$ is defined in Eq \eqref{eq:errordelta1}.

Since $O\phi_\theta = (O_1'+O_2' +\delta O)\phi_\theta = 0$, we find
\begin{equation}
\begin{split}
    &|\tr[(O_1'+O_2')^\dagger (O_1'+O_2') \sigma_\theta]| = 2|\tr[(O_1'+O_2')^\dagger \delta O\sigma_\theta]| \\
    & \leq \const\cdot L^{3d-1-\alpha+a}.
\end{split}
\end{equation}
For $\alpha>3d-1$, we can always choose $a$ and $\beta$ such that this expression vanishes for large $L$. Hence,
\begin{equation}
    \begin{split}
      F := 2\tr(O_1'^\dagger O_1'\sigma_\theta) + \tr(O_1'^\dagger O_2' \sigma_\theta) + \tr(O_2'^\dagger O_1' \sigma_\theta) = 0,  
    \end{split}
\end{equation}
where we utilize $\tr(O_1'^\dagger O_1'\sigma_\theta) = \tr(O_2'^\dagger O_2'\sigma_\theta)$, up to a vanishing error again for $\alpha>3d-1$. Due to the LR bound, $\int_0^{L^\beta} d\tau A(i\tau)-\hc$ and $\delta_1$ can be approximated by an operator acting within $l'$ and $u'$, respectively.  Employing this approximation gives an error of the order specified in Eq \eqref{eq:operatorrestriction} in the quantity $F$ (or Eq \eqref{eq:operatorrestrictionklocal} for generic $k$-local Hamiltonians), which vanishes for large $L$ when $\alpha>3d-2$. Consequently, we can substitute $T_{\theta,-\theta}$ by $T_{\theta,0}$ in $O_1'$, since now the operator conjugated by $T_{\theta,-\theta}$ acts as an identity near coordinate $x=\frac{L}{2}$ (and similarly substitute $T_{\theta,-\theta}$ by $T_{0,-\theta}$ in $O_2'$). Thus, we deduce that
\begin{equation}
    F = 2\tr(O_{l'}^\dagger O_{l'}\sigma_\theta) + \tr(O_{l'}^\dagger O_{u'} \sigma_\theta) + \tr(O_{u'}^\dagger O_{l'} \sigma_\theta) = 0,
\end{equation}
where $O_{u'}$ is the restriction of $O_2'$ in $u'$.

Finally, let us consider the \emph{connected} correlation function
\begin{equation}
    C_{l'u'} = \tr(O_{l'}^\dagger O_{u'} \sigma_\theta) - \tr(O_{l'}^\dagger  \sigma_\theta)\tr( O_{u'} \sigma_\theta).
\end{equation}
Since $O_l'$ and $O_u'$ are supported within $l'$ and $u'$ respectively, their distance $r$ is at least $\frac{L}{8}$. It has been demonstrated that if $\sigma_\theta$ is the unique gapped ground state of a power-law Hamiltonian, the connected correlation between two operators (with unit norm) separated by $r$ is upper bounded by a function $G(r)$ decaying in $r$ with the same rate as the LR bound of Eq \eqref{eq:LRboundform} \cite{2020hierarchy}, i.e., $G(r)\simeq\frac{1}{r^{\alpha-d}}$.\footnote{In the expression for $G(r)$, the volumes of the regions $u'$ and $l'$ do not appear due to $S_z^{(0)}$, $S_z^{(\frac{L}{2})}$, $\partial_\theta H_\theta^{tw}$, and $\partial_\theta H_\theta^u$ being sums of operators acting on a finite number of sites. Specifically, up to a vanishing error quantified by Eq.\eqref{eq:operatorrestriction}, $C_{l'u'}\simeq \langle O_1'^\dagger O_2'\rangle_c = \tr(O_1'^\dagger O_2' \sigma_\theta) - \tr(O_1'^\dagger  \sigma_\theta)\tr( O_2' \sigma_\theta)$. Let us consider the contribution $\langle \int_0^{L^\beta}d\tau [A^+(i\tau)-\hc] \delta_1\rangle_c$ to $\langle O_1'^\dagger O_2'\rangle_c$, while other terms can be analyzed similarly. We have:
\begin{equation}
    \begin{split}
        \langle \int_0^{L^\beta}d\tau [A^+(i\tau)-\hc] \delta_1\rangle_c = -\frac{1}{\pi^2} \int_0^{L^\beta} d\tau_1 d\tau_2 e^{-\frac{\Delta^2(\tau_1^2+\tau_2^2)}{2q}} \\
        \times\int_{t_1,t_2}\frac{t_1 t_2 e^{-\frac{\Delta^2(t_1^2+t_2^2)}{2q}}}{(t_1^2+\tau_1^2)(t_2^2+\tau_2^2)} \langle \partial_\theta H^{tw}_\theta (t_1-t_2) \partial_\theta H_\theta^u \rangle_c.
    \end{split}
\end{equation}
As $\partial_\theta H_\theta^u$ is a sum of terms, each acting on \emph{two} spins within $u$, and $H^{tw}_\theta (t_1-t_2)$ can be approximated by an operator within $l'$, we thus have $\langle \partial_\theta H^{tw}_\theta (t_1-t_2) \partial_\theta H_\theta^u \rangle_c\leq \frac{\const}{L^{\alpha-d}}\Vert \partial_\theta H^{tw}_\theta \Vert\cdot \Vert \partial_\theta H^u_\theta \Vert$. This immediately leads to Eq. \eqref{eq:connectedc}.} We therefore conclude that
\begin{equation}
\begin{split}
    C_{l'u'}\leq \const \cdot L^{2(d-1+\frac{a}{2})-\alpha+d},
    \end{split}
    \label{eq:connectedc}
\end{equation}
Here in the calculation we use the fact that, for $\alpha>d+1$, the dominant contribution to $\int_0^{L^\beta} d\tau [A^{+,-}(i\tau)-\hc]$ and $\delta_1$ again comes from spin-spin couplings in the vicinity of the two twists, at links $(L-1,0)$ and $(\frac{L}{2}-1,\frac{L}{2})$, respectively, \ie Eq \eqref{eq:areatwist}. For $\alpha>3d-2$, we can choose $a$ and $\beta$ such that this expression vanishes for large $L$.

Lastly, we claim that
\begin{equation}
    \tr(O_{l'}^\dagger  \sigma_\theta) = \tr( O_{u'} \sigma_\theta) =0.
    \label{eq:onepointfunction}
\end{equation}
This can be easily seen by noting that $\tr(O_1'^\dagger \sigma_\theta)$ and $\tr(O_2' \sigma_\theta)$ vanish. This is because $S_z$, $\partial_\theta H_\theta^u|_{\theta=0}$, and $\partial_\theta H_\theta^l|_{\theta=0}$ are odd under a spin rotation of $\pi$ along the $S_x$ direction, and the assumption that $|\phi_0\rangle$ is $SO(3)$ symmetric. Given that $O_{l'}$ and $O_{u'}$ are spatial truncations of $O_1'$ and $O_2'$ respectively, Eq \eqref{eq:onepointfunction} holds up to an error of the order specified in Eq \eqref{eq:eaerror}. Consequently, we have demonstrated that $\tr(O_{l'}^\dagger O_{l'} \sigma_\theta)$ vanishes for large $L$, thus concluding the proof of the theorem. For generic $k$-local Hamiltonians, Theorem \ref{thm:higherd} holds when the parameter $\alpha$ governing the decay of the interaction in Eq. \eqref{eq:fastdecay} satisfies $\alpha > \max(3d, 4d - 2)$.

\emph{Type II Hamiltonians}

We conclude this paper by discussing Type II Hamiltonians. We construct the twisted Hamiltonian in a similar manner. However, a crucial distinction arises. In a Type I Hamiltonian $H_I = \sum_{ij} h_{ij}$, following the conjugation $H_\theta = U_R^\dagger(\theta) H_I U_R(\theta)$, all terms supported simultaneously on the left and right halves undergo modification. The number of modified terms amounts to approximately $L^{2d}$. In contrast, for a Type II Hamiltonian $H_{II} = \sum_{ij} \frac{O_i O_j}{(|i-j|+1)^\alpha}$, only those terms such that either $O_i$ or $O_j$ act exactly on the twist, i.e., across the links $(L-1,0)$ or $(\frac{L}{2}-1,\frac{L}{2})$, are modified, totaling approximately $L^{2d-1}$. This observation similarly extends to $H_\theta^u$ and $H_\theta^l$, as well as to any generic $k$-local Hamiltonians.

Following a similar line of reasoning, we construct a twisted state as in Eq.\eqref{eq:twisted} and examine the reduced density matrices in $\Bar{u}$ (the region complementary to the upper quarter), denoted as $\rho_\theta^1$. Again, we aim to demonstrate that
\begin{equation}
\begin{split}
\partial_\theta \rho_\theta^1 - \partial_\theta \sigma_\theta^1
= e_\theta = e_a + e_b
\end{split}
\label{eq:1errorII}
\end{equation}
has a small trace norm, where the two errors $e_a$ and $e_b$ are defined in Eq.\eqref{eq:twoerrors}. Repeating the derivation for Type I Hamiltonians, one can demonstrate that the magnitude of $e_a$ remains unchanged, as indicated in Eq.\eqref{eq:klocalea}. This is because for all generic $k$-local Hamiltonians satisfying the two requirements in Eq.\eqref{eq:fastdecay} and Eq.\eqref{eq:extensive}, the LR bound in Eq.\eqref{elseLR} holds, and this bound straightforwardly applies to Type II Hamiltonians, which are a special subclass of $k$-local Hamiltonians.

On the other hand, the estimation for $e_b$ is slightly different. Repeating the derivation, we again claim that in the first term of $e_b$ (the commutator), substituting $A^{+,-}$ with $B^{+,-}$ introduces only a small error. Let us evaluate the error $\delta$ resulting from this substitution:
\begin{equation}
    \begin{split}
        \delta = & -\frac{1}{2\pi i} \int_0^{L^\beta} d\tau e^{-(\Delta\tau)^2/2q}\\ 
        & \times\int_t \frac{[\partial_\theta H^{tw}_\theta-\partial_\theta H_\theta](t)}{t+i\tau} e^{-(\Delta t)^2/2q}-\hc.
        \end{split}
        \label{eq:errorreplaceII}
\end{equation}
A crucial observation is that $\partial_\theta H^{tw}_\theta-\partial_\theta H_\theta$, and therefore $\delta$, can be divided into two categories, $\delta = \delta_1 + \delta_2$: (1) $\delta_1$ arises from terms supported entirely within $u$, as defined in Eq \eqref{eq:errordelta1}; (2) $\delta_2$ comes from all remaining terms in $\partial_\theta H^{tw}_\theta-\partial_\theta H_\theta$, denoted as $H_m$. Notably, each term in $H_m$ has a diameter of at least $\frac{L}{8}$. Crucially, for a generic $k-$local Type II Hamiltonian, $H_m$ contains only terms acting non-trivially on the twist. Using Eq. \eqref{eq:fastdecay}, we find that $\Vert H_m \Vert$ is bounded by $O(\frac{1}{L^{\alpha-2d+1}})$. The contribution from terms in the second category to $\delta$ is then upper bounded by Shanti's bound:
\begin{equation}
\begin{split}
    \Vert \delta_2  \Vert  \leq  \frac{\Vert H_m \Vert}{2} \sqrt{\frac{2\pi q}{\Delta^2}} \leq \const\cdot L^{2d-1-\alpha+\frac{a}{2}}.
\end{split}
\label{eq:delta2II}
\end{equation}
Compared with Eq. \eqref{eq:delta2}, we find that for the special subclass of Type II Hamiltonians, $\delta_2$ becomes smaller. This explains why the applicability of the theorem is extended.

We now consider $\delta_1$. Following the derivation leading to Eq. \eqref{eq:finaldelta1}, we need to bound the operator norm of $[O,\delta_1]$, where $O$ is an operator supported within $\Bar{u}$ with unit norm. Note that $\partial_\theta H_\theta^u$ can be further divided into two parts, $\partial_\theta H_\theta^u = K_1 + K_2$: (1) The terms $K_1$ that are supported solely in the regions $k_1 \cup k_2$ (as depicted in Fig. \ref{Fig:system1}). Note that the minimal distance from each of these terms to $O$ is $\frac{L}{16}$. Following exactly the same argument leading to Eq. \eqref{klocalerrorea}, for a generic $k$-local Hamiltonian with properties \eqref{eq:fastdecay} and \eqref{eq:extensive}, the LR bound constrains the commutator
\begin{equation}
   \begin{split}
       &\Vert [-\frac{1}{2\pi i} \int_0^{L^\beta} d\tau e^{-(\Delta\tau)^2/2q}\int_t \frac{K_1(t)}{t+i\tau} e^{-(\Delta t)^2/2q}-\hc,O] \Vert \\
       & \leq\const\cdot L^{\frac{a}{2}(\frac{d}{1-\sigma}+1)+\min(\beta,\frac{a}{2})+(1-\sigma)(\alpha-d)-(\alpha-2d+1)}.
   \end{split}
   \label{eq:delta1K1II}
\end{equation}

(2) $K_2$ includes all other terms in $K$, namely, terms that act non-trivially in the region $k_3$ or $k_4$, or both. Note that the diameter of each term in $K_2$ is at least $\frac{L}{16}$, and therefore $\Vert K_2 \Vert \leq O(L^{2d-1-\alpha})$. As a result, we have (Shanti's bound)
\begin{equation}
    \begin{split}
       &\Vert [-\frac{1}{2\pi i} \int_0^{L^\beta} d\tau e^{-(\Delta\tau)^2/2q}\int_t \frac{K_2(t)}{t+i\tau} e^{-(\Delta t)^2/2q}-\hc,O] \Vert \\
       & \leq\const\cdot L^{2d-1-\alpha+\frac{a}{2}}.
   \end{split}
   \label{eq:delta1K2II}
\end{equation}
Compared with Eq \eqref{eq:delta1K2}, for the special class of Type II Hamiltonians, this error becomes smaller due to the fact that they only contain couplings among $SO(3)$ singlet local operators, and therefore $K_2$ contains only terms that act non-trivially on the twist.

Combining Eqs \eqref{eq:delta2II}, \eqref{eq:delta1K1II} and \eqref{eq:delta1K2II}, and following the same derivation as for a generic Type I Hamiltonian, we arrive at
\begin{equation}
\begin{split}
&\Vert e_\theta \Vert_1\simeq \Vert e_\theta' \Vert_1 \\
\leq & O( L^{\frac{a}{2}(\frac{d}{1-\sigma}+1)+\min(\beta,\frac{a}{2}) + (1-\sigma)(\alpha-d) - (\alpha-2d+1)}).
\end{split}
\label{errorthetaII}
\end{equation}
As a result, for $\alpha>2d$ we have
\begin{equation}
    \tr(H_\theta^{tw} \rho_\theta) \leq \const \cdot L^d ( \Vert e_\theta \Vert_1 + \Vert e_\theta' \Vert_1 ) +O(L^{2d-\alpha}),
\end{equation}
where we utilized the requirement Eq. \eqref{eq:extensive}, which ensures that the energy is extensive, \ie $\Vert H_\theta^{tw} \Vert \simeq O(L^d)$. Thus, for $\alpha > 3d-1$, it is always possible to select suitable values for $\sigma$, $a$, and $\beta$ such that the energy of $\rho_{2\pi}$ vanishes with $L$ algebraically. This completes the construction of a low-lying twisted state.

It can also be shown that the lattice momentum of the twisted state $\psi_{2\pi}$ is $\pi$, making it orthogonal to the original ground state $\phi_0$ when $\alpha>4d-3$. The derivation closely resembles that of the Type I Hamiltonian scenario, with the only distinction being the magnitude of Eq. \eqref{eq:Holder}, now given by
\begin{equation}
\begin{split}
& 2\pi \Vert O_{l'} \Vert^2 \cdot \Vert e_\theta \Vert_1 \leq  \const \Vert O_1 \Vert^2 \cdot \Vert e_\theta \Vert_1 \\
\leq & \const\cdot L^{4d-3-\alpha+\min(\beta,\frac{a}{2})+\frac{a}{2}(\frac{d}{1-\sigma}+3)+(1-\sigma)(\alpha-d)},
\end{split}
\end{equation}
since now the magnitude of $\Vert e_\theta \Vert_1$ is given by Eq. \eqref{errorthetaII}. For any $\alpha > 4d-3$, it is possible to select $\sigma$ sufficiently close to 1 and $a > \beta > 0$, both sufficiently small, ensuring the vanishing of this error in the thermodynamic limit. We conclude that in $d$ spatial dimensions, a $k-$local Type II Hamiltonian cannot have a unique gapped ground state for $\alpha>\max(3d-1,4d-3)$.


\end{document}